\documentclass{article}
\usepackage{spconf,amsmath,graphicx}
\usepackage{booktabs}
\usepackage{color,amsmath,url,times, tabularx,bbm,amssymb,multirow}
\usepackage{hyperref}
\usepackage{microtype}
\usepackage{cite}
\usepackage{graphicx}
\usepackage{subcaption, subcaption}
\usepackage[para]{footmisc}
\usepackage{booktabs} %
\usepackage{comment}
\usepackage{makecell}
\usepackage{boldline}
\usepackage{amsfonts,amsmath,amssymb, amsthm} 
\usepackage{mathtools}
\usepackage{multirow}
\usepackage{microtype}
\usepackage{xspace}
\usepackage[ruled, vlined]{algorithm2e}
\usepackage{siunitx}
\DeclareSIUnit{\params}{\relax}
\usepackage{balance}
\newcommand{\pluseq}{\mathrel{{+}{=}}}
\newcommand{\minuseq}{\mathrel{{-}{=}}}
\usepackage[table]{xcolor}%
\newcommand{\PreserveBackslash}[1]{\let\temp=\\#1\let\\=\temp}
\newcolumntype{C}[1]{>{\PreserveBackslash\centering}p{#1}}

\usepackage{hyperref}
\usepackage{array}
\usepackage{tikz}
\newcolumntype{H}{>{\setbox0=\hbox\bgroup}c<{\egroup}@{}}

\title{HiFi-Codec: Group-residual Vector quantization for High Fidelity Audio Codec}

\name{Dongchao Yang$^{1*}$, Songxiang Liu $^{2*}$, Rongjie Huang$^{3*}$, Jinchuan Tian$^{1}$, Chao Weng $^{2}$, Yuexian Zou$^{1}$
      \thanks{Dongchao Yang, Songxiang Liu and Rongjie Huang are the main contributor for this project. Work done during an internship at Tencent AI Lab}
\address{$^{1}$ Peking University, China\\
        $^{2}$ Tencent AI Lab, Shenzhen, China \\
        $^{3}$ Zhejiang University, China \\
}}
\begin{document}
%\ninept
%
\maketitle
\begin{abstract}
%is a effective metric-based few-shot learning method
Audio codec models are widely used in audio communication as a crucial technique for compressing audio into discrete representations. Nowadays, audio codec models are increasingly utilized in generation fields as intermediate representations. For instance, AudioLM is an audio generation model that uses the discrete representation of SoundStream as a training target, while VALL-E employs the Encodec model as an intermediate feature to aid TTS tasks. Despite their usefulness, two challenges persist: (1) training these audio codec models can be difficult due to the lack of publicly available training processes and the need for large-scale data and GPUs; (2) achieving good reconstruction performance requires many codebooks, which increases the burden on generation models. In this study, we propose a group-residual vector quantization (GRVQ) technique and use it to develop a novel \textbf{Hi}gh \textbf{Fi}delity Audio Codec model, HiFi-Codec, which only requires 4 codebooks. We train all the models using publicly available TTS data such as LibriTTS, VCTK, AISHELL, and more, with a total duration of over 1000 hours, using 8 GPUs. Our experimental results show that HiFi-Codec outperforms Encodec in terms of reconstruction performance despite requiring only 4 codebooks. To facilitate research in audio codec and generation, we introduce AcademiCodec, the first open-source audio codec toolkit that offers training codes and pre-trained models for Encodec, SoundStream, and HiFi-Codec.
Code and pre-trained model can be found on: \href{https://github.com/yangdongchao/AcademiCodec}{https://github.com/yangdongchao/AcademiCodec}
\end{abstract}
\section{Introduction}
\label{sec:intro}
The purpose of an audio codec is to reduce the amount of data needed to store or transmit an audio signal without significantly degrading its quality. The basic principle behind audio codecs is to remove redundant or irrelevant information from the audio signal. There are many different types of audio codecs, each with its own strengths and weaknesses. Some codecs are designed for use in real-time applications, such as telephony or streaming audio, and prioritize low latency and low bitrates. Others are designed for high-quality audio production and prioritize fidelity and accuracy. In addition to compression, many audio codecs also include features such as error correction, noise reduction, and dynamic range compression. These features can help to improve the quality and reliability of the audio signal, especially in challenging environments such as noisy or low-bandwidth networks.

In this study, we focus on using audio codec models to help solve generation problems in audio-related works, such as Text-to-Speech (TTS)\cite{wang2023neural,yang2023instructtts,huang2022fastdiff,huang2022prodiff}, music generation \cite{agostinelli2023musiclm}, audio generation \cite{yang2022diffsound,kreuk2022audiogen,borsos2022audiolm,huang2023make,huang2022transpeech}. The Encodec \cite{defossez2022high} and SoundStream \cite{zeghidour2021soundstream} are the most related to our work.
The core technique in Encodec and SoundStream is residual vector quantization (RVQ), which uses multiple VQ codebooks to represent the intermediate features. Encodec and SoundStream both adopt an encoder-decoder framework. The encoder first compresses the waveform into compact deep representations, then RVQ is used to quantize the intermediate features. Lastly, the decoder is used to recover waveform from the quantized representations. 

In our experiments, we found that the most of information is saved in the first codebook when RVQ is used, \textit{e.g.} for speech data, text information, and timbre information can be recovered by only using one codebook. The following codebooks save some details information, which may influence the audio quality. However, such information is sparse and scattered in the hidden space. Thus, such a quantization style needs many codebooks to realize good reconstruction performance. \textit{e.g.} Encodec needs 12 codebooks to realize high-quality reconstruction performance. In audio generation fields, using multiple codebooks will bring burden to the generation model, \textit{e.g.} long sequence is hard to model by the transformer.  

In this study, we focus on designing an audio codec model for generation tasks. Our goals are mainly two-fold: 1) good reconstruction performance, and 2) require a small number of codebooks. To realize this, we propose group-residual vector quantization (GRVQ) methods. For any latent features $\boldsymbol{z} \in \mathbb{R}^N$, we first split $\boldsymbol{z}$ into several group, \textit{e.g.} \{$\boldsymbol{z}_1$, $\boldsymbol{z}_2 $\}. Then we use residual vector quantization (RVQ) to quantize $\boldsymbol{z}_1$ and $\boldsymbol{z}_2$, respectively. Lastly, we combine the information from two groups of RVQ to decode the waveform. Our motivation is that we expect the first layer's codebook can save more information so that we can use fewer residual blocks. When we split the features into several groups and use more codebooks in the first layer, these codebooks in the first layer can play more important factors in the compression process.  In our experiments, we found that we split the features into 2 groups, and using 2 residual layers can bring good reconstruction performance than pre-trained Encodec model \footnote{https://github.com/facebookresearch/encodec}.
\section{Related works}
\subsection{Audio Representation Learning.}
The usage of self-supervised learning (SSL) \cite{baevski2020wav2vec,baevski2022data2vec} has got great success in audio-related fields, such as auto-speech recognition (ASR) and audio compression \cite{zeghidour2021soundstream, defossez2022high}. 
Inspired by vector quantization (VQ) techniques \cite{van2017neural},
SoundStream \cite{zeghidour2021soundstream} presents a residual vector quantization (RVQ) architecture for high-level representations that carry semantic information. Similarly, many works \cite{iashin2021taming, yang2022diffsound} try to use VQ-VAE model to compress the time-frequency spectrogram (\textit{e.g.} Mel-spectrogram) into high-level representations.
\subsection{Speech and Audio Generation with Audio Codec}
Recently, many works \cite{wang2023neural, yang2023instructtts,agostinelli2023musiclm,borsos2022audiolm,kreuk2022audiogen,chen2023vector,huang2023audiogpt} propose to model speech and audio in the discrete latent space with the help of Audio Codec. The core idea is that using an audio codec model compresses the speech or sound into a group of discrete tokens, then uses a generation model to generate these tokens. \textit{e.g.} AudioLM \cite{borsos2022audiolm} utilize the audio codec model encodes the waveform into discrete tokens, and then uses Language Model (LM) to model the generation process. Similarly, InstructTTS \cite{yang2023instructtts} uses discrete diffusion models to generate discrete tokens, then uses an audio codec model to recover the waveform. 

\subsection{Audio Codec}
The study of low-bitrate parametric audio codecs dates back to \cite{atal1971speech, juang1982multiple}, but their quality is often limited. Recently, researchers have proposed several neural network-based audio codecs that show promising results \cite{kleijn2018wavenet,valin2019real, zeghidour2021soundstream, omran2022disentangling, jayashankar2022architecture,defossez2022high}. These methods typically use an encoder to extract deep features in a latent space, which is then quantized before being fed to the decoder.
The most relevant related works to ours are the SoundStream  \cite{zeghidour2021soundstream} and Encodec \cite{defossez2022high} models, where these methods propose to use a fully convolutional encoder-decoder architecture with a Residual Vector Quantization (RVQ) \cite{gray1984vector, vasuki2006review} layers. These models were optimized using both reconstruction loss and adversarial perceptual losses. To accelerate the process of compression and decompression, Encodec proposes a language model method to predict tokens. MQ-TTS \cite{chen2023vector} also proposes to use multiple codebooks to quantize intermediate features, but they assume speaker information can be explicit from the speaker labels and do not use residual VQ to preserve more audio information. Although previous works have got great success in terms of reconstruction performance and compression rate, these works may not be suitable for generation tasks due to many codebooks being needed to maintain good reconstruction performance.
\begin{algorithm}[t]
\DontPrintSemicolon
\caption{Group-Residual Vector Quantization}\label{algo:grvq}
\SetKwInput{Input}{Input}
\SetKwInput{Output}{Output}
 \Input{ $\boldsymbol{y} = encoder(x)$ the output of the encoder, vector quantizers $Q_i$ for $i=1..2*N_q$} 
\Output{ the quantized $\boldsymbol{\hat{y}}$ }
$\hat{y}_1 \gets 0.0$\;
$\hat{y}_2 \gets 0.0$\;
split $\boldsymbol{y}$ into two group; \\
${\rm residual}_1 \gets \boldsymbol{y}_1$\;
${\rm residual}_2 \gets \boldsymbol{y}_2$\;
\For{$i=1$ to $N_q$}{
$\hat{y}_1 \pluseq Q_i({\rm residual}_1)$\;
${\rm residual}_1 \minuseq Q_i({\rm residual}_1)$\;
}
\For{$i=N_q$ to $2*N_q$}{
$\hat{y}_2 \pluseq Q_i({\rm residual}_2)$\;
${\rm residual}_2 \minuseq Q_i({\rm residual}_2)$\;
}
$\boldsymbol{\hat{y}} = concat(\boldsymbol{\hat{y}_1},\boldsymbol{\hat{y}}_2)$\\
\textbf{return} $\boldsymbol{\hat{y}}$
\end{algorithm}
\begin{figure*}[t]
  \centering
  \includegraphics[width=\linewidth]{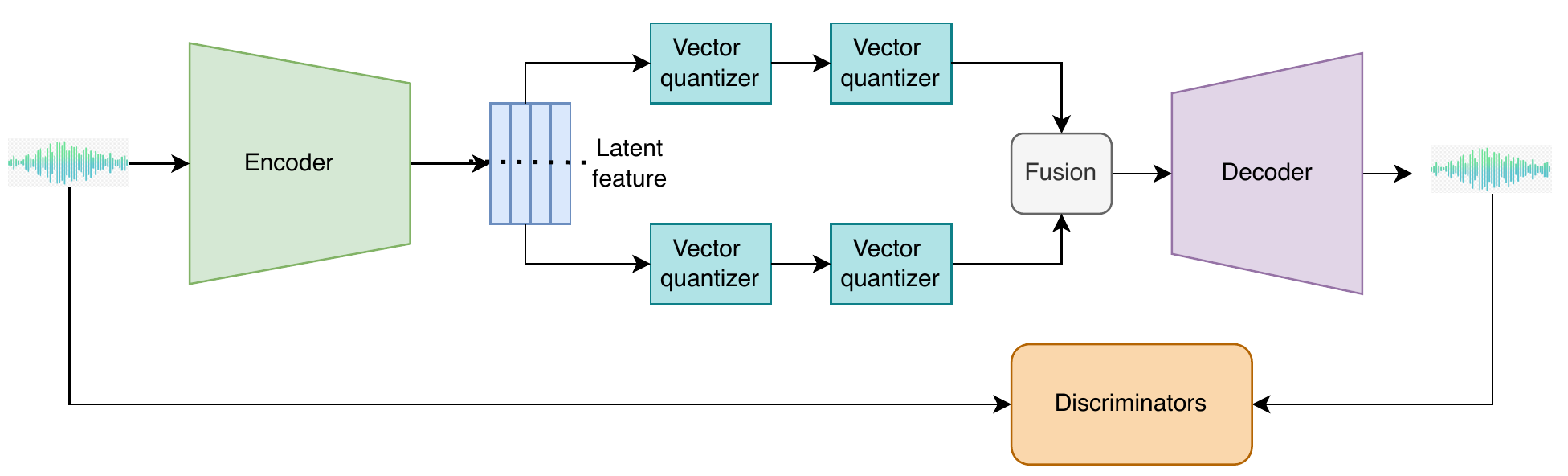}
  \caption{The overview of HiFi-Codec model.}
  \label{fig:ml}
%   \vspace*{-\baselineskip}
\end{figure*}
\section{Proposed Method}
In this section, we will introduce the details of HiFi-Codec. We first introduce the overview of HiFi-Codec model, then we discuss the details of each part in HiFi-Codec.
\subsection{Overview}
We consider a single-channel audio signal $\boldsymbol{x}$ with duration $d$, represented as a sequence $\boldsymbol{x} \in \mathcal{R}^{T}$, where $T=d*sr$ and $sr$ is the audio sample rate. The HiFi-Codec model comprises three main components: (1) an encoder network $E$ that takes the input audio and generates a latent feature representation $\boldsymbol{z}$; (2) a group-residual quantization layer $Q$ that produces a compressed representation $\boldsymbol{z}_q$; and (3) a decoder $G$ that reconstructs the audio signal $\boldsymbol{\hat{x}}$ from the compressed latent representation $\boldsymbol{z}_q$. The model is trained end-to-end, optimizing a reconstruction loss applied over both time and frequency domains, along with a perceptual loss in the form of discriminators operating at different resolutions. A visual description of the proposed method can be seen in Figure 1. 
\subsection{Encoder and Decoder}
Our model's encoder and decoder architecture draws inspiration from the designs of Encodec \cite{defossez2022high} and SoundStream \cite{zeghidour2021soundstream}. The architecture is based on a convolutional framework with sequential modeling applied over the latent representation on both the encoder and decoder sides.
The encoder model $E$ comprises a 1D convolution with $C$ channels and a kernel size of 7, followed by $B$ convolution blocks. Each convolution block features a single residual unit, which is followed by a down-sampling layer consisting of a strided convolution with a kernel size $K$ of twice the stride $S$. The residual unit consists of two convolutions with a kernel size of 3 and a skip connection. The number of channels is doubled whenever down-sampling occurs. The convolution blocks are then followed by a two-layer LSTM for sequence modeling and a final 1D convolution layer with a kernel size of 7 and $D$ output channels.
In our study, we explore different settings for $C$, $B$, and $S$, such as $C = [32, 48, 64]$, $B = 4$, and $S=(2,4,5,8)$ or (2, 4, 5, 6) or (2, 2, 2, 4). The decoder mirrors the encoder and uses transposed convolutions instead of stride convolutions, with the strides in reverse order as in the encoder. The decoder outputs the final audio signal.
\subsection{Group-residual Vector Quantization (GRVQ)}
In this study, we want to design an audio codec model that contains fewer quantizers while enjoying good reconstruction performance. We think that one of the drawbacks of the RVQ is that the first layer of codebooks in RVQ will save the most of information, but the remaining codebooks only save a little information. Thus we propose to add more codebooks in the first layer. Specifically, for any latent feature representation $\boldsymbol{z}$, we first split into several groups averagely (in our study, we split $\boldsymbol{z}$ into two groups, $\boldsymbol{z}_1, \boldsymbol{z}_2$), and using multiple RVQ to quantize each group features. Lastly, we combine multiple group RVQ output to obtain the final quantization results. The whole process can be summarized as Algorithm 1.
\begin{table*}[t] \centering
\caption{The performance comparison.}
\label{tab:my-table}
\begin{tabular}{cccccc}
\hline
Method             & Sample rate (K hz) & Down-sample times & Number of codebooks & PESQ $\uparrow$  & STOI $\uparrow$  \\ \hline
Encodec (Facebook) & 24                & 320         & 8         & 3.01 & 0.94 \\ 
Encodec (Facebook) & 24                & 320         & 12         & 3.21 & 0.95 \\ \hline
Encodec (ours)     & 24              & 240         & 8         & 3.62  & 0.94 \\
Encodec (ours)     & 24               & 32          & 2         & 3.08  & 0.91  \\
Encodec (ours)     & 16               & 320         & 8         & 3.04  & 0.93  \\ \hline
SoundStream (ours)     & 16               & 320         & 12         & 3.26  & 0.95  \\ \hline
HiFi-Codec       & 24               & 240         & 4         & 3.63  & 0.95 \\
HiFi-Codec       & 24               & 240         & 8         & \textbf{3.92}  & \textbf{0.95} \\
HiFi-Codec       & 24               & 320         & 4         & 3.64  & 0.95 \\ 
HiFi-Codec       & 16               & 320         & 4         & 3.22  & 0.94 \\ \hline
\end{tabular}
\end{table*}
\subsection{Discriminator}
In this study, we use three discriminators: A multi-scale STFT-based (MS-STFT) discriminator, which is used on Encodec \cite{defossez2022high}; a multi-period discriminator (MPD) and a multi-scale discriminator (MSD) from HiFi-GAN vocoder \cite{kong2020hifi}. \\
For the MS-STFT discriminator, which consists in identically structured networks operating on multi-scaled complex-valued STFT with the real and imaginary parts concatenated. We adopt the same configuration as Encodec for each sub-network, which consists of a 2D convolutional layer followed by 2D convolutions with increasing dilation rates in the time dimension (1, 2, and 4) and a stride of 2 over the frequency axis. A final 2D convolution with a kernel size of 3 x 3 and stride (1, 1) provides the final prediction. We use five different scales with STFT window lengths of [2048, 1024, 512, 256, 128].
For the multi-period discriminator and multi-scale discriminator, we maintain the same structure as HiFi-GAN, but reduce the channel number to make the discriminator have similar parameters to MS-STFT.
\subsection{Training Loss}
Our approach is based on a GAN objective, in which we optimize both the generator and the discriminators. Specifically, we jointly optimize a reconstruction loss term, a perceptual loss term (via discriminators), and the GRVQ commitment loss for the generator. The training objective of the generator comprises several loss terms, including a time domain term, a frequency domain term, three discriminator losses, and the corresponding feature loss terms acting as a perceptual loss and the GRVQ commitment loss. The discriminator loss is based on the adversarial hinge-loss function.
\subsubsection{Reconstruction Loss}
Our reconstruction loss comprises two aspects: (1) time domain loss and (2) time-frequency loss. For the time domain loss, we directly use the L1 distance loss to optimize $\boldsymbol{x}$ and $\boldsymbol{\hat{x}}$. For the time-frequency loss, we follow a similar approach to Encodec and apply a loss term on the mel-spectrogram with several time scales.
\subsubsection{Discriminator loss}
The adversarial loss is used to promote perceptual quality. We use three types of discriminator, where MS-STFT discriminator try to make the spectrogram-level reconstruction results as similar as the original one. MPD and MSD discriminators try to make the waveform-level reconstruction results as similar as the original one. To train the discriminator, we can optimize the following objective function:
\begin{align}\label{dis loss}
    \mathcal{L}_{\mathit{d}} = \frac{1}{K} \sum_{i=1}^K \mathit{max}(0, 1-D_k(\boldsymbol{x}))+\mathit{max}(0,1+D_k(\boldsymbol{\hat{x}})) 
\end{align}
where $K$ denotes the number of discriminators. Furthermore, we 
can define the adversarial loss as a hinge loss over the logits of these discriminators:
\begin{align}\label{adv loss}
    \mathcal{L}_{\mathit{adv}} = \frac{1}{K} \sum_{i=1}^K \mathit{max}(0, 1-D_k(\boldsymbol{\hat{x}}))
\end{align}
Furthermore, the feature loss is computed by taking the average absolute difference between the discriminator's internal layer outputs for the generated audio and those for the corresponding real audio.
\begin{align}\label{vqvae loss}
    \mathcal{L}_{\mathit{feat}} = \frac{1}{KL} \sum_{k=1}^K \sum_{l=1}^L \frac{||D_k^l(\boldsymbol{x})-D_k^l(\boldsymbol{\hat{x}})||_1}{\mathit{mean}(||D_k^l(\boldsymbol{x})||_1)}
\end{align}
\subsubsection{GRVQ Commitment Loss}
For the i-th group c-th residual quantizer, we can calculate the commitment loss based on following formula:
\begin{align}\label{vqvae loss}
    \mathcal{L}_{\mathit{c}} = \sum_{i,c} ||\boldsymbol{z}_{i,c}-q_{i,c}(\boldsymbol{z}_{i,c})||_2^2 
\end{align}
Based on previous discussion, we can use following formula to train the generator.
\begin{equation}
    Loss_G = \lambda_{\text{adv}}  \mathcal{L}_{\mathit{adv}} + \lambda_{\text{feat}} \cdot \mathcal{L}_{\mathit{feat}} + \lambda_{\text{rec}} \cdot   \mathcal{L}_{\mathit{rec}} + \lambda_c \cdot  \mathcal{L}_{\mathit{c}}
\end{equation}
where  $\mathcal{L}_{\mathit{adv}}$ denotes the adversarial loss. $\mathcal{L}_{\mathit{feat}}$ denotes the feature loss. $\mathcal{L}_{\mathit{rec}}$ denotes the reconstruction loss. $\lambda_{\text{adv}}$, $\lambda_{\text{feat}}$, $\lambda_{\text{rec}}$ and $\lambda_c$ are the hyper-parameters to control the training objective function. In our experiments, we try to balance each loss terms by scale these hyper-parameters. 
\section{Experiments}
\subsection{Evaluation metric}
In this study, we evaluate the audio codec model's performance by measuring the gap between reconstruction audio and the target one. We adopt the metrics from speech enhancement fields, such as the PESQ and STOI to evaluate the performance.
\subsection{Dataset}
We use TTS dataset to train audio codec models. Our training data comes from public datasets, such as LibriTTS, VCTK, AISHELL, which mainly includes English and Chinese speech.
\subsection{Experimental results}
Table 1 shows the experimental results. We can see that our proposed HiFi-Codec realizes good reconstruction performance while only using 4 codebooks. The best performance can be abtained when we set downsample times as 240, and the number of codebooks as 8 (we set the each layer includes 4 codebooks, and two residual layers are used). Furthermore, our reproduced models (Encodec and SoundStream) also get comparable performance with Encodec \cite{defossez2022high}. We strongly recommend readers to use the HiFi-Codec model with 4 codebooks when readers try to train a generation model.

\section{Conclusion}
In this study, we present group-residual vector quantization method, and build a novel audio codec model: HiFi-Codec, which is specially designed for generation tasks. HiFi-Codec can bring better reconstruction performance than Encodec even using 4 codebooks. Furthermore, we also release the training process of Encodec and SoundStream models, which can help readers to train their own codec models. In the future, we will continue to optimize the HiFi-Codec models, and try to train better Encodec and SoundStream models. We expect this project can facilitate the research in audio generation tasks.
\section{Limitations}
Although HiFi-Codec models realize good construction performance than Encodec model, the limitations still exist. (1) We do not use large-scale dataset to train a universal audio codec, the generalization cannot be validated very well. (2) We find that the objective evaluation metrics may not very accurate to assess the reconstruction performance. Subjective evaluation is always the best choice, but this part is missed in this study. (3) HiFi-Codec aims to help generation tasks, but we donot provide enough down-stream tasks to evaluate the performance. We take this direction to our future works.

\bibliographystyle{IEEE.bst}
\bibliography{refs.bib}

\end{document}